\documentclass[
    ,final            
  ]
  {aipproc}

\layoutstyle{6x9}

\begin{document}

\title{Fueling and morphology of central starbursts}

\author{Johan H. Knapen}{
address={Centre for Astrophysics Research, University of Hertfordshire,
Hatfield, Herts AL10 9AB, U.K.},
}

\begin{abstract}

Of the mechanisms proposed to bring gaseous fuel into the central
starburst regions of a galaxy, non-axisymmetries in the gravitational
potential set up by interactions or by bars are among the most
promising. Nevertheless, direct observational evidence for a
connection between interactions and bars on the one hand, and
starburst (as well as AGN) activity on the other, remains
patchy. These general issues are reviewed before proceeding to
discuss, within this context, massive star formation in the
circumnuclear regions of spiral galaxies, in particular star-forming
nuclear rings and pseudo-rings. Such rings are common, and are closely
linked to the dynamics of their host galaxy, which in almost all cases
is barred.  The possible existence of a population of nuclear rings on
scales which are too small for detection with standard ground- or even
space-based techniques will be discussed.

\end{abstract}

\maketitle


\section{Introduction}

Some of the most important open questions relating to starburst
activity in galaxies are what combination of physical conditions leads
to the triggering of the burst, and how long a starburst can be
sustained. These questions are intimately related to another one,
namely how a starburst is fueled. One obvious prerequisite for the
occurrence of a starburst is the availability of sufficient gaseous
fuel for the much enhanced star formation, of the right properties,
e.g., temperature, density, and dynamics, and at the right location at
the right time.

Important parallels exist between the problems of fueling starbursts
and active galactic nuclei (AGN), at least low-level AGN such as
Seyferts and LINERs. Both will need a well-regulated supply of gas
being fed into the nuclear region of a galaxy from its disk, and both
can be expected to be correlated with agents capable of driving such
inflow. Of these, bars and galaxy interactions are the most important,
because they lead to non-axisymmetries in the gravitational potential
which in turn lead to gas shocks and subsequent loss of angular
momentum, allowing the gas to move radially inward.

In this review paper, the current observational evidence for fueling
starbursts by means of non-axisymmetries in the galactic potential,
specifically bars and interactions, is briefly reviewed, making
reference to parallel work on Seyfert host galaxies.  Various aspects
of nuclear rings are then considered, which can be seen as a specific
morphological class of low luminosity starbursts. Relations between
nuclear rings and their host galaxies are explored, as well as
possible links between the occurrence of galactic rings and the
presence of nuclear starburst or AGN activity. The size distribution
of nuclear rings is also addressed. The main points addressed and
conclusions reached in this review are outlined briefly in the final
section.

\section{Fueling of starbursts}

The so-called ``fueling problem'' in starbursts (as well as in AGN) is
not the amount of gas present, but how to transport this gas from the
main body of the host galaxy to the central region where it is needed
to feed the activity. Assuming typical starburst gas consumption rates
and lifetimes (of, say, a couple of solar masses per year and of order
$10^7-10^8$ years) it is easily seen that the outer regions of the
host galaxies of starbursts will in general contain more than enough
gas to fuel the burst. The problem, then, is how to deliver this gas
to the starburst region in the center of the galaxy. To transport the
gas radially inward, it must lose most of its angular momentum, for
which a number of mechanisms can be invoked. The most important of
these are gravitational, driven by a non-axisymmetry in the galactic
potential set up by a bar or a galaxy interaction or merger, and
effective on spatial scales of tens of parsecs to kiloparsecs, and
possibly even smaller than that.

There is an important parallel here between the fueling of central
starbursts and the fueling of AGN, both of which require similar gas
delivery methodology into the nuclear regions of the host galaxy. In
AGN, the problem may be more acute because the scales on which
material needs to be delivered are most probably much smaller than in
the case of starbursts, possibly on the scale of AUs ($\sim10^{-6}$
parsec) for AGN. The fueling of starbursts and especially AGN has been
reviewed rather extensively in the recent literature, and here only a
brief summary of the work relevant to gravitational fueling of
starbursts will be given. For more complete discussions, although
mostly biased to the fueling of AGN rather than starbursts, the reader
is referred to reviews by Shlosman, Begelman \& Frank (1990), Combes
(2001), Shlosman (2003), Knapen (2004a,b), and Jogee (2004).

\subsection{Bars fueling starbursts}

Bars are expected theoretically and numerically to lead to gas
concentration toward the central regions of galaxies because gas in
the bar can lose angular momentum due to torques and shocks (e.g.,
Schwarz 1984; Combes \& Gerin 1985; Noguchi 1988; Shlosman, Frank \&
Begelman 1989; Knapen et al. 1995a,b). There is some direct
observational evidence that bars indeed instigate central
concentration of gas, the most comprehensive of which comes from
surveys of molecular gas concentration in barred as compared to
non-barred galaxies, as traced through the emission by CO
molecules. Sakamoto et al. (1999) found that, statistically, the ten
barred galaxies in their sample have more concentrated CO emission,
and thus, presumably, molecular hydrogen, than their ten non-barred
galaxies (see also Sheth et al. 2002 for a confirmation of this result
with larger samples). The reader is referred to Knapen (2004a) for a
more detailed discussion of the evidence for gas concentration by
bars, as well as of some of the caveats which need to be taken into
account.

Accepting that bars cause gas inflow, the question is then whether
this also leads to AGN or central starburst activity. On the former,
the bar fraction is indeed higher among Seyfert host galaxies than
among non-Seyferts, but the statistical significance of this result
does not quite reach 3$\sigma$. Moreover, the bar fractions (of 80\%
and 60\%, respectively) indicate that most galaxies are barred,
irrespective of their nuclear activity, and that there are significant
numbers of galaxies, even Seyfert hosts, in which no bar can be
distinguished (Knapen, Shlosman \& Peletier 2000; Laine et al. 2002;
Laurikainen, Salo \& Buta 2004; see Knapen 2004b for a more detailed
review).

Nuclear starbursts clearly occur preferentially in barred hosts (e.g.,
Hummel 1981; Hawarden et al. 1986; Devereux 1987; Dressel 1988;
Puxley, Hawarden, \& Mountain 1988; Arsenault 1989; Huang et al. 1996;
Martinet \& Friedli 1997; Hunt \& Malkan 1999; Roussel et
al. 2001). Huang et al. (1996) studied IRAS data and confirmed that
starburst hosts are preferentially barred, although with the proviso
that the result holds for strong bars (SB in RC3 terminology) and in
early-type galaxies only, as later confirmed by Roussel et al. (2001).
All studies referred to above use optical images or information from
catalogs such as the RC3 (de Vaucouleurs et al. 1991) to derive
whether a galaxy is barred or not. Given the higher bar fractions and
higher accuracy in bar determination that are achieved when using
near-IR images, there is thus considerable scope for further study, by
using near-IR images of well-selected samples of starburst and control
galaxies.

\subsection{Interactions fueling starbursts}

Since interactions between galaxies can lead to angular momentum loss
of inflowing material and thus to gas inflow, they can be invoked to
explain the fueling of starbursts and AGN (e.g., Shlosman et al. 1989,
1990; Mihos \& Hernquist 1995). For Seyfert galaxies, however, there
is {\it no evidence at all} in the literature for statistical
connections between galaxy mergers and interactions on the one hand,
and the occurrence of AGN activity on the other (see Knapen 2004b for
a review with relevant references to other work). And unlike the case
of bars which has been discussed above, there is a marked difference
here between the behavior of low luminosity AGN and starburst
activity, with the latter very clearly influenced by interactions, at
least in the extremes. This is most obvious in the case of the
ultra-luminous infrared galaxies (ULIRGs), powered mainly by very
powerful starbursts (Genzel et al.  1998) and, apparently without
exception, associated with galaxies involved in mergers or other
strong interaction processes (e.g., Joseph \& Wright 1985; Armus,
Heckman, \& Miley 1987; Sanders et al. 1988; Clements et al. 1996;
Murphy et al. 1996; Sanders \& Mirabel 1996). It is clear that the
most extreme of the starbursts need the occurrence of the
destructively powerful event in the host galaxy which is a galaxy
merger, but it is unlikely that all such mergers lead to the kind of
starbursts exemplified by the ULIRGs (or even more normal, lower
luminosity starbursts). It is not even clear whether interactions lead
to enhancements of the star formation activity {\it in general}.
Illustrating this point are the results from Bergvall, Aalto \&
Laurikainen (2003), who find no evidence for significantly enhanced
star-forming activity among interacting/merging galaxies as compared
to non-interacting galaxies, but do report a moderate increase in star
formation in the very centers of their interacting sample galaxies.

\section{Nuclear rings}

Nuclear rings are common, occurring in around one fifth of all disk
galaxies (Knapen 2005). Their properties have been reviewed in detail
by Buta \& Combes (1996), with more recent aspects described by
Kormendy \& Kennicutt (2004) and Knapen (2004a, b). Galactic rings,
including nuclear ones, can form in the vicinity of resonances in a
gaseous disk, usually set up by a bar or other form of non-axisymmetry
in the gravitational potential. Gas concentrates near such resonances,
can become unstable, and collapse to lead to the massive star
formation associated with rings. Nuclear rings are associated with
Inner Lindblad Resonances (ILRs), usually occur in barred galaxies,
and have typical radii of $0.5-1$~kpc, issues which are discussed in
more detail, below. Nuclear rings form significant numbers of stars
which can help shape a pseudo-bulge, and so drive secular evolution
(Kormendy \& Kennicutt 2004), but which also contribute a couple of
percent to the overall star formation rate of the local universe
(R.~C.~Kennicutt, these proceedings, p. 000).

\subsection{Nuclear rings and their host galaxies}

Nuclear rings have long been known to occur preferentially in barred
galaxies, and in early-type disk galaxies (see review by Buta \&
Combes 1996). Knapen (2005) quantified this using H$\alpha$ imaging
of a statistically representative sample of 57 spiral
galaxies\footnote{These data are part of a larger data set, consisting
of images in the $B, R, I, K_{\rm s}$ broad bands, as well as in
H$\alpha$, of the 57 sample spirals. The description of these data has
been published by Knapen et al. (2003, 2004a), and the images
themselves are freely available to interested researchers through the
Centre de Donn\'ees Stellaires via anonymous ftp to
cdsarc.u-strasbg.fr (130.79.128.5) or via
http://cdsweb.u-strasbg.fr/cgi-bin/qcat?J/A+A/426/1135, or through the
NASA/IPAC Extragalactic Database (NED).}. The distribution of the
morphological types of the nuclear ring host galaxies peaks at type
Sbc, while no rings were found in types later than Scd, and very few
in earlier types. In later types, the bulge may not be massive enough
to ensure the presence of ILRs, whereas in earlier types, there may
not be enough gas present.

Of the 12 nuclear rings in Knapen's (2005) survey, 10 occur in
galaxies classified as either SAB or SB in the RC3, but two are
classified as non-barred. Upon closer inspection, though, it turns out
that both these galaxies, NGC~1068 and NGC~4736, do have bars, albeit
rather small, and only visible in near-IR imaging (Scoville et
al. 1988; Thronson et al.  1989; Shaw et al.  1993; M\"ollenhoff et
al.  1995; Wong \& Blitz 2000; Laine et al. 2002; Knapen 2005). This
is a general picture in the literature, where by far most nuclear
rings are associated with bars in their host galaxies. There are,
however, a number of cases where galaxies without any trace of a bar,
even in the near-IR, do show prominent nuclear rings. As an example,
careful imaging in optical and near-IR bands, from both the ground and
from the {\it Hubble Space Telescope (HST)}, confirms the absence of a
bar in the disk of the small nuclear ring host galaxy NGC~278 (even
though it has been classified as SAB in the RC3). However, 21cm H{\sc
i} imaging reveals that in the outer regions of this small galaxy both
the gas morphology and kinematics are severely distorted, which
indicates that NGC~278 has recently undergone a minor merger event
with an even smaller galaxy, possibly similar to a Magellanic Cloud,
which would have led to a destabilization and non-axisymmetry of the
gravitational potential, and hence to gas inflow, the formation of
resonances, and ultimately the formation of the star-forming nuclear
ring (Knapen et al. 2004b). It is not known at present whether other
unbarred nuclear ring host galaxies, such as NGC~7742, may exhibit the
same clues to an interactive past from H{\sc i} data, but this would
be well worth checking.

\begin{figure}[ht]
\includegraphics[height=.35\textheight]{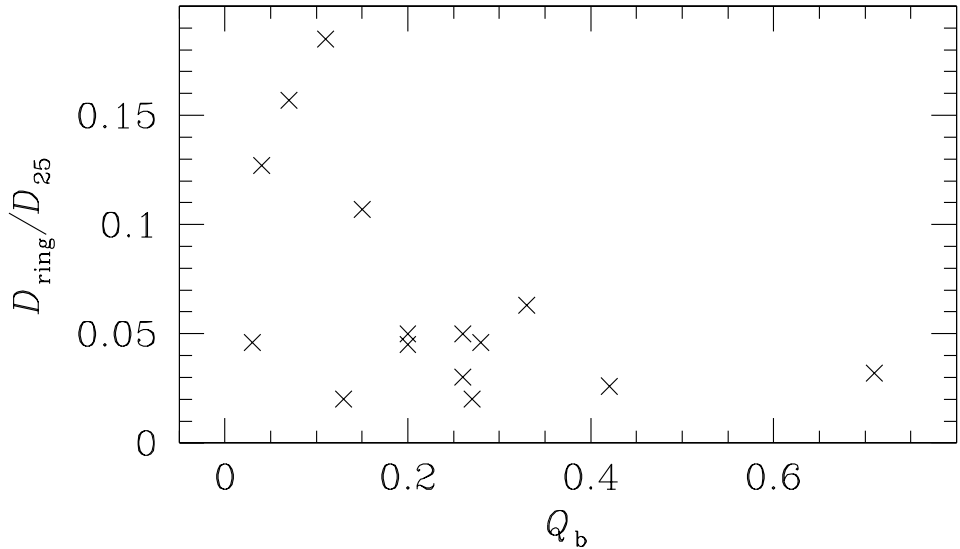}
\caption{Relative size,  or ring diameter divided  by galaxy diameter,
of 15 nuclear rings as a function of the gravitational torque ($Q_{\rm
b}$) of the bar of its host galaxy. Data from Knapen (2005).}
\end{figure}

In general, though, we can say that observed ring statistics (reviewed
by Buta \& Combes 1996) firmly support a picture in which resonances
set up by a galactic bar induce the formation of rings. Moreover, we
can also confirm that, as expected by numerical modeling and theory,
the properties of the bars shape the properties of the nuclear
ring. As an example, we mention the relation between bar strength and
nuclear ring size, as reported by Knapen, P\'erez-Ram\'\i rez \& Laine
in 2002, and recently confirmed with a larger number of rings by
Knapen (2005). We found that whereas nuclear rings of all sizes can
and do occur in weak bars, there is a complete absence of {\it large}
nuclear rings from galaxies with {\it strong} bars\footnote{Here, we
use the ratio of ring and host galaxy radius as a measure of nuclear
ring size, and the gravitational torque parameter, $Q_{\rm b}$, as a
measure of the bar strength (for the latter see the work by Buta,
Block \& Knapen 2003 and Block et al.  2004, and references therein to
earlier work).}  (Fig.~1). This result nicely agrees with theoretical
expectations, in which the ILRs, and thus the nuclear rings, occur
near the interface between the parallel (to the bar) $x1$ and
perpendicular $x2$ orbit families. Since rings depend on gas
concentration, they can only occur where the spatial extent of these
two orbit families does not overlap.  In strong bars, which have
larger ellipticity (are ``thinner'') than weak bars, the physical
space around the galaxy nucleus where non-overlapping $x2$ orbits can
occur is smaller, and as a result only smaller rings can form in such
bars (see Heller \& Shlosman 1996).  The absence of large rings in
strong bars is thus strong support for the picture where nuclear rings
are formed as a result of resonances set up by a galactic bar.

\subsection{Galactic rings and nuclear activity}

The possible causal connections between galactic rings and nuclear
activity have received relatively little attention in the literature,
as compared to the amount of attention received as separate topics, or
even in comparison to topics like the connection between nuclear
activity and other host galaxy properties such as the presence of bars
or interactions. Anecdotally, a number of famous galaxies host both an
AGN and a nuclear ring or pseudo-ring, for example, NGC~1068 or
NGC~4303. The relative lack of attention may be at least partly due to
the fact that one might not {\it a priori} expect relationships
between rings and nuclear activity. The spatial scales of even the
smallest nuclear rings are much larger than those of AGN activity, and
the timescales of the two phenomena are essentially unknown but not
necessarily related. In addition, rings are a known consequence of the
presence of bars in disks, and logically bars have received more
attention as possible primary instigators of nuclear activity. For
starburst activity such a link has indeed reliably been found, but for
AGN activity the link is not nearly as direct (see the discussion
earlier in this paper and in the references listed there).

\begin{figure}[ht]
\includegraphics[height=.5\textheight]{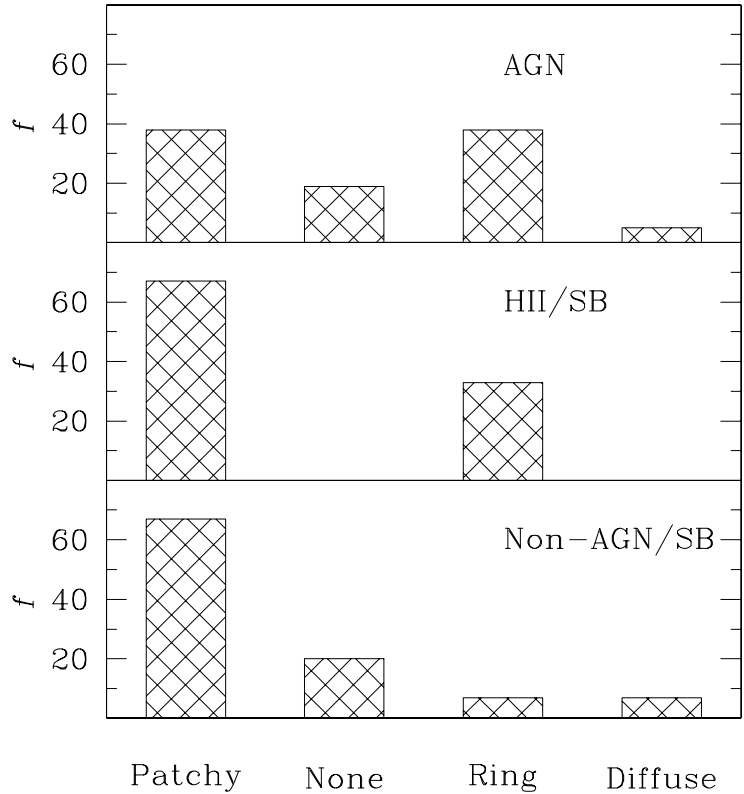}
\caption{Distribution of nuclear H$\alpha$ emission morphology in those
galaxies classified as AGN (including both Seyfert and LINER) in the
NED ({\it top panel}), those classified as starburst or H{\sc ii}
({\it middle panel}), and those galaxies classified as neither ({\it
lower panel}).  Fractions are given as percentages of the total
numbers of galaxies in each category (AGN, etc). Adapted from Knapen
(2005).}
\end{figure}

There is some statistical evidence from the literature which indicate
the potential of rings as tracers of the fueling of nuclear
activity. For instance, Arsenault (1989) reported a higher incidence
of the combination of bar and ring among starburst and AGN hosts than
among ``normal'' galaxies, while Hunt \& Malkan (1999) found that
LINERs and Seyfert galaxies have significantly more rings (inner and
outer) than normal galaxies or starbursts. In both these studies, the
RC3 morphological classification was used to derive bar and ring
frequencies. More recently, Knapen (2005) correlated the presence of
a nuclear ring as identified from H$\alpha$ imaging with the presence
of nuclear activity (both starburst and AGN). The latter was obtained
from the NED, but the use of more robustly defined activity indicators
(e.g., from Ho, Filippenko \& Sargent 1997), not available for all
sample galaxies, would not change the results significantly.  The
results are surprising: not only is the circumnuclear H$\alpha$
emission morphology of the AGN and starbursts in the sample of 57
galaxies significantly more often in the form of a nuclear ring than
in non-AGN (only 7\% of non-starburst, non-AGN and 11\% of non-AGN
galaxies have circumnuclear rings, versus 33\% of starbursts and 38\%
of AGN), but also nuclear rings occur significantly more often than
not in galaxies which also host nuclear activity (while 27 of the 57
sample galaxies are starburst or AGN hosts, 10 of the 12 nuclear rings
occur in this class of galaxy; see Fig.~2).

There are, thus, some indications for a statistical connection between
galactic rings and the occurrence of nuclear activity. For starburst
activity this would not be surprising, since galaxies with relatively
compact nuclear rings with a high star formation rate might well be
classified as starburst, whereas on the other hand a starburst
classified as nuclear may in fact be circumnuclear, but not resolved
with imaging. For Seyfert and LINER activity, such a link is more
puzzling, even though the connection between that type of activity and
starbursts is now well known (e.g., Cid Fernandes et al. 2004 and
references therein). Especially for the outer and inner rings, but
also for nuclear rings which have typical radii of a kiloparsec, the
spatial scales are very different from those of the AGN (thought to be
as small as AU-scale), whereas the timescales of both the AGN activity
and the star-forming phase in the rings may well be shorter than the
dynamical time needed to bridge the range in spatial scales. For bars,
especially when invoking bars within bars, all these problems seem
less prominent than for rings, so it is puzzling that the statistical
connections between rings and AGN, if confirmed, are more
significant. All these aspects of rings and nuclear activity are in
urgent need of further scrutiny.

\subsection{Are small nuclear rings common?}

\begin{figure}[ht]
\includegraphics[height=.4\textheight]{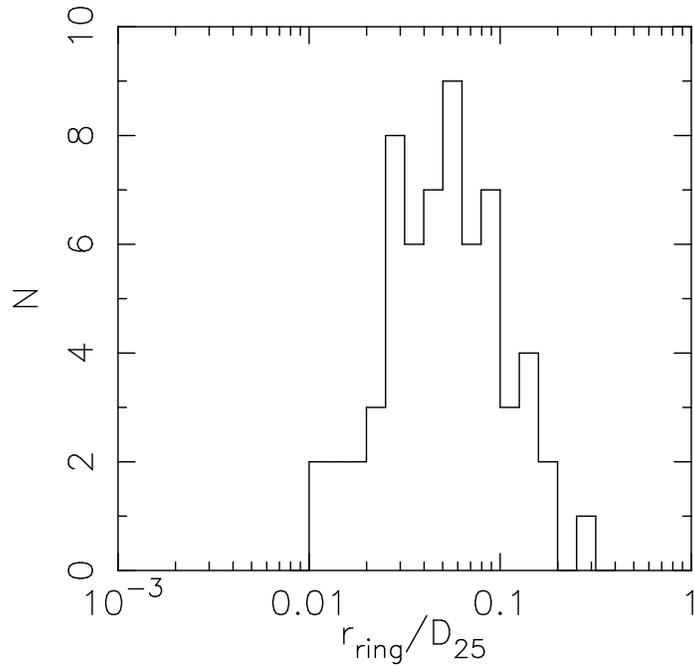}
\caption{Distribution of the relative sizes of 62 nuclear rings 
(ring size divided by galaxy size). Reproduced with permission from
Laine et al. (2002).}
\end{figure}

A fair number of nuclear rings are now known in the literature. Laine
et al.  (2002) compiled sizes for 62 of them, mostly from Buta \&
Crocker (1993).  Laine et al. plotted the size distribution of these
rings, and found a marked peak in ring size with a maximum near
$r_{\rm ring}/D_{25}=0.06$ (Fig.~3).  The size distribution is rather
narrowly peaked, coinciding with the typical radius of ILRs in the
galaxies under consideration. It is possible, though, that the cutoff
in the distribution on the low end is due to observational bias, where
the smaller rings have simply not been resolved. Knapen (2005) points
out that the median distance to galaxies hosting a nuclear ring in his
sample of 57 (12.1~Mpc) is significantly smaller than the median
distance to other groups of galaxies in his sample, or in fact the
whole sample of 57 (16.9~Mpc), but even so, rings with sizes
significantly below the peak value in Fig.~3 should be rather easily
observable. It thus seems that there is a true absence of small
nuclear rings.

There is anecdotal evidence in the literature, however, for nuclear
rings which are significantly smaller than the typical 1~kpc for
nuclear rings. For instance, in NGC~5248, Laine et al. (2001) and Maoz
et al. (2001) found a ring of some 0.1~kpc (two~arcsec) in radius
(Jogee et al. 2002), NGC~278 contains a small ring of 0.2~kpc radius
(about four arcsec; Knapen et al. 2004b), and IC~342 hosts a
nuclear ring of less than 100~pc in diameter (B\"oker,
F\"orster-Schreiber, \& Genzel 1997). So can there be a significant
population of nuclear rings on spatial scales small enough to have
escaped detection in nuclear ring survey such as those by, e.g., Buta
\& Crocker (1993) or Knapen (2005)?

There are two systematic studies which can shine light on this
issue. In the first of these, Maoz et al. (1996a)  only found one
hitherto undetected nuclear ring, which at 4~arcsec diameter should be
observable from the ground.  Maoz et al. used UV imaging from the {\it
HST} with a spatial resolution of $\sim0.05$~arcsec. The overall
fraction of nuclear rings reported by Maoz et al. is low ($\sim5\%$),
primarily because the field of view of their images is only 22~arcsec
(as recognized by Maoz et al. 1996b), implying that six or seven of
the 12 nuclear rings of Knapen (2005) would have fallen outside their
imaged areas, but also because the effects of dust extinction are
pronounced at the UV wavelengths used by Maoz et al., even though the
effects of the latter on nuclear ring detection are not clear. The
latter issue also precludes one from concluding that small nuclear
rings do not exist. The smaller the radius of the nuclear ring, the
more pronounced the effects of dust extinction will be, hence the more
difficult to recognize nuclear rings from UV imaging. 

The second study, by B\"oker et al. (1999) is based upon a series of
{\it HST} NICMOS Pa$\alpha$ images of the central (51~arcsec across)
regions 94 spirals, at a spatial resolution of 0.2 arcsec. In their
study, five galaxies only have been indicated to host, or possibly
host, a nuclear ring. Inspection of the published images reveals that
only four are {\it bona fide} rings, of which two are inner rings, and
only those in NGC~2903 and NGC~1241 can be considered nuclear
rings. The one in NGC~2903 is well-known from ground-based images
(e.g., Wynn-Williams \& Becklin 1985; see also Alonso-Herrero \&
Knapen 2001), the one in NGC~1241 has a diameter of four arcsec and is
thus also large enough to be observed from the ground,as has in fact
been done by, e.g., Mazzuca et al. (these proceedings,
p.~000). Interestingly, with respect to the discussion on possible
relationships between nuclear activity and rings, all four ring hosts
from the survey of B\"oker et al.  are AGN or starburst hosts!

Although it thus seems unlikely that there is a significant population
of small nuclear rings (with radii of around tens of parsecs or
smaller), further imaging data of well-selected samples of galaxies,
and at high resolution (ideally higher than the 0.2~arcsec of B\"oker
et al. (1999), must be analysed in detail to definitely settle this
issue.

\section{Concluding remarks} 

The main points discussed in this short review are the following:

\begin{itemize}

\item Bars and interactions, because of the deviations from symmetry 
they cause in the gravitational potential of the host galaxy, are the
most appropriate mechanisms to consider as fueling agents, which serve
to remove angular momentum from gas and thus allow it to move radially
inward in galaxies. Both starbursts and AGN will depend on the
continued supply of such gaseous material to sustain their activity.

\item It is clear that the most extreme starbursts, exemplified by the 
ULIRGs, almost exclusively occur in strongly interacting
galaxies. Whether in general interactions lead to starbursts, or
whether starbursts need an interaction to be triggered, are issues
which are less clear. There are no indications that interactions are
related to the occurrence of Seyfert- or LINER-type non-stellar
activity in galaxies.

\item Bars are statistically linked to starbursts, possibly with 
stronger connections in certain subclasses such as strong bars and
early-type hosts. Seyfert hosts are also more often barred than
non-Seyferts, but at a lower significance level.

\item Nuclear rings are common, occurring in at least 
one in every five disk galaxies. Their host galaxies are generally of
morphological type around Sbc, and are almost without exception
barred. Nuclear rings are low-luminosity starbursts, and are commonly
believed to occur as a result of massive star formation in gas
accumulated near ILRs, and are thus resonance phenomena whose
properties can be used to chart the dynamics of the host galaxy.

\item Several studies report a possible statistical connection 
between the presence of rings in a galaxy and the presence of nuclear
activity. Although the statistics and/or the data quality of these
works should be improved, it is noteworthy that the ring-AGN relation
may well turn out to be more prominent than the bar-AGN
relation. Although the simultaneous presence of starburst and AGN
activity in a galaxy has now been firmly established as a rather
common occurrence, any link between rings, even nuclear ones, and
non-stellar activity is puzzling because the spatial scales involved
are different by many orders of magnitude. Further scrutiny, first
observationally and subsequently phenomenologically, is needed.

\item Small nuclear rings, on scales of tens or hundreds of parsecs, 
are known in a handful of well-studied nearby galaxies, but evidence
for the presence of a significant population of such rings is so far
lacking.

\end{itemize}

{\bf Acknowledgements} I thank my collaborators on the various
research projects described in this paper, especially Isaac Shlosman
who also provided comments on an earlier version of this manuscript.

\clearpage

\end{document}